# Taking Turing by surprise?
# Designing 'digital computers' for morally-loaded contexts

Sylvie Delacroix[1]

*University of Birmingham*

## Abstract

There is much to learn from what Turing hastily dismissed as Lady Lovelace's 'objection': 'digital computers' can indeed surprise us. Just like a piece of art, algorithms can be designed in such a way as to lead us to question our understanding of the world, or our place within it. Some humans do lose the capacity to be surprised in that way: it might be fear, or it might be the comfort of ideological certainties. As lazy normative animals, we do need to be able to rely on authorities to simplify our practical reasoning: that's ok. Yet the growing sophistication of computer systems designed to free us from the constraints of normative engagement may well take us past a point of no-return: what if, through lack of normative exercise, our 'moral muscles' became so atrophied as to leave us unable to question our social practices?

This paper makes two distinct normative claims:

1. Decision-support systems should be designed with a view to regularly jolting us out of our moral torpor.
2. Without the depth of habit to somatically anchor model certainty, a computer's experience of something new cannot but remain very different from that which in humans gives rise to non-trivial surprises. This asymmetry has important repercussions when it comes to the shape of ethical agency in 'artificial moral agents': the worry is not just that they would be likely to leap morally ahead of us, unencumbered by the weight of habits. The main reason to doubt that the moral trajectories of humans v. autonomous systems might remain compatible stems from the asymmetry in the mechanisms underlying moral change. Whereas in humans surprises will continue to play an important role in waking us to the need for moral change, cognitive processes will rule when it comes to machines. This asymmetry cannot but translate into increasingly different moral outlooks, to the point of (likely) unintelligibility. The latter prospect is enough to doubt the desirability of autonomous moral agents.

---

[1] I would like to thank Chris Baber, Simon Blackburn, Mireille Hidebrandt, Jurgen Van Gael, Michael Veale, and Alan Wilson for their helpful comments on an earlier draft. I also benefited from excellent feedback from the computer science work-in-progress seminar here in Birmingham. The research leading to this work was funded by the Leverhulme Trust.

# Table of Contents

We have devised machines that help us keep track of time, cultivate the earth, mend our bodies, survey the skies…The list goes on. Some aim to overcome specific physical limitations; others are designed to entertain. Many do both. Most have had a profound impact on our understanding of the world, and the role we can play within it, none more so than one of our recent inventions: 'digital computers'.

In that context, to ask whether computers are able to 'take us by surprise' may sound like a redundant question. When it is famously raised by Alan Turing (Turing, 1950), nobody is in a position to predict the depth and extent of the social, cultural, and intellectual upheavals brought about by their near-universal use today. Yet this historical upheaval is not what Turing has in mind when he floats the 'computers cannot take us by surprise' proposition only to better dismiss it as unsubstantiated. Turing indeed points out that computers do take him by surprise all the time, given their ability to fill the gaps in his incomplete (or deficient) calculations.

That reply is a bit quick, and not only because it hinges upon some lasting ambiguity in the meaning of 'surprise'. One can indeed use the word quite prosaically, to refer to any event or 'input' that we fail to anticipate, whether or not it has any impact on our understanding of ourselves, the world, or our place within it. In that mundane sense, we cannot but be surprised all the time, given our very limited ability to store data and process it. Yet that prosaic sense does not quite capture the context within which Turing raises his 'surprise challenge'. Turing indeed does so as an attempt to translate the so-called 'Lovelace Objection' to his claim that computers will ultimately be able to fool an external observer whose aim is to tell the computer from the human in a question and answer game. Among the arguments that may be raised to dispute this claim, one may point at various 'disabilities' of computers. Lady Lovelace alludes to one of them when, describing Babbage's 'Analytical Engine' (in 1842), she notes that the latter 'has no pretension to *originate* anything. It can do *whatever we know how to order it* to perform'.[2]

---

[2] Quoted by (Turing, 1950, p. 12)

Lady Lovelace's objection is important in at least two respects. First, negatively, because it spells out a common (and mistaken) assumption according to which true autonomy presupposes an ability to initiate something in a way that is unprecedented, in the sense that it is not conditioned by or relying on prior norms. The romantic appeal of such a notion of unprecedented beginning as a way of conceptualising our human ability (and responsibility) to give ourselves norms has misled many. Modern history has been scarred by it: the French revolutionaries' perceived need to start a new calendar may be the most telling illustration of the dire political consequences that can flow from such an assumption.[3] When it comes to beginnings, and our ability to give ourselves norms, we actually have a fair bit in common with computers: we do not start from scratch. We develop our norms on the basis of prior norms and expectations. Woven into the fabric of social interaction that structures our daily lives, these prior norms are both given to us and re-formulated by us on a continuous basis – just like the norms that preside over some computers' ability to learn from their environment (under a reinforcement learning model).

Yet Lady Lovelace's objection also hints at an important way in which 'digital computers'[4] may never be like us: for want of the deeply internalised patterns of behaviour (or thought) that constitute human habits, their experience of something new is unlike that which in humans gives rises to 'surprises' in the non-prosaic sense. Given the importance of this thesis for the rest of this paper, the whole of section 1 is dedicated to it: while section 1.1. explores the notion of surprise – as a mechanism underlying model change – within the Machine Learning literature, section 1.2, analyses the conceptual link between non-trivial surprises and habit. Without the depth of habit to emotionally (and somatically) 'anchor' model certainty, a computer's experience of something new cannot but remain very different from that which in humans gives rise to non-trivial surprises.

---

[3] In *The Human Condition*, Arendt draws an important distinction, based on St Augustine, between 'the beginning which is man (*initium*)' and 'the beginning of the world (*principium*)' (Arendt, 1998). According to Arendt, if the French revolutionaries had not understood their task as an absolute, godlike beginning (*principium*, beginning of the world), which is by definition beyond their capacities, they would probably have been able to avoid many of the perplexities (and ensuing violence) they were confronted with.

[4] The `digital' qualification is important, for the word `computer' can be used to characterise humans too: more on this in Section 1. In the rest of this paper, `computer' is used a short for `digital computer'.

This asymmetry has important consequences when it comes to ethical agency and the shape it would take in so called 'autonomous moral agents': it is not just that the latter would be likely to 'leap morally ahead' of us, unencumbered by the weight of habits (if it were, preserving mutual intelligibility would merely be a matter of decelerating their evolution). It is also that their moral trajectory would likely be qualitatively different, given the fundamental difference in the mechanisms underlying moral change. The implications of this asymmetry are spelt out in section 2, which focuses on the problematic assumptions underlying current efforts to ethically 'train' (or constrain) autonomous systems that are capable of being deployed in morally loaded contexts.

Having highlighted the role played by both habit and surprise within the mechanisms underlying moral change in humans, section 3 considers the consequences of this conceptual link when it comes to designing decision-support systems meant for morally loaded contexts. These decision-support systems should be designed in a way that fosters - rather than dulls- the situational awareness without which humans lose the ability to be surprised in a non-trivial sense. Why? Because all too often, all that stands between us and the worst atrocities is best described as the surprise that stems from encountering 'the Other'. In Levinas' work, this encounter lies at the root of our experiencing ethical responsibility, by reminding us of our common humanity, and the inescapable demands that stem from it. Rather than shielding us from such encounters, decision-support systems can and should be designed to facilitate them. This paper concludes with the following question: as significantly *different* 'Others', might computers ever be 'encountered' by us in a way which gives rise to non-trivial, ethically significant surprises too? I believe so, and in that way hope that Lady Lovelace will be proven wrong, even if this paper aims to show the importance of the insights underlying what Turing too quickly dismissed as her 'objection'.

# 1. Experiencing novelty: asymmetry between humans and computers

To ask whether computers can 'originate' anything in a way that surprises us (as per Lady Lovelace's objection), one must presuppose a notion of 'surprise' that goes beyond the mere unanticipated event or proposition: neither humans nor algorithms are capable of 100% accurate predictions in real-life applications. Prediction failures are expected. Some events or propositions (whether they were anticipated or not) do nevertheless give rise to surprises in a way that goes beyond the trivial, 'unanticipated' sense; they do so when they lead us to reconsider our understanding of ourselves, the world, or our role within it.

Some humans do lose the capacity to be surprised in that way. The force of habit(s) and/or the comfort of our 'certainties' can all too easily prevent the un-anticipated from ever disconcerting us, with dire consequences in terms of ethical agency. Section 3 explains why (and how) computers should be designed to help us in that respect. To understand what is at stake, however, one needs to grasp the asymmetry between the mere 'unanticipated even/or proposition' -whose experience is within any computer's reach (1.1)- and the existential questioning of one's understanding of the world (or one's place within it) that stems from a genuinely surprising event or proposition. The depth of that experience is conceptually linked to the mechanisms underlying habit formation (and change) in humans (1.2).

## 1.1. 'Surprises' and machine-learning optimisation strategies

Our ability to be surprised in a non-trivial sense is most commonly compromised by habit and/or certainties. To what extent, if any, do computers face the same hurdles? The answer to that question is largely dependent upon what one means by 'computers': it is positive if one refers to those 'persons who carried out calculations', as in its original use. It is currently negative if one refers to 'digital computers' or 'Turing machines' (henceforth: 'computers'). It is indeed far from clear whether such machines could ever internalise repeated patterns of behaviour in a way that gives rise to habits *as humans experience* *them*. The latter qualification (in italics) is important: to be efficient, algorithms do capitalise on repeated patterns (via optimisation or `bundling' strategies). The difference -and it is key- consists in the nature of the effort needed to break from such patterns. This is discussed in section 2.1. As for certainties, alogirthms can be designed so as to preserve the fallibility of whatever model they rely on to go about their daily tasks.

Recent research in fact draws attention to a problem commonly associated with Bayesian learning methods: because posterior uncertainty is reduced with each step – no matter how 'surprising' the data sample may be – model uncertainty reaches close to zero as the number of data samples increases. The resulting 'inability to be surprised' compromises the learning capacity of the system. The latter indeed requires a balance to be found between the plasticity necessary to being able to draw upon new knowledge-generating experiences and the stability without which learned memories get forgotten. This insight is far from new.[5] Yet its re-discovery within the field of Machine Learning is giving rise to some fascinating surprise-focused research[6], including attempts at 'quantifying' surprises[7] so as to devise mathematical tools to 'help' Machine Learning applications retain the ability to be surprised. These tools are designed to ensure that 'a small model uncertainty remains even after a long stationary period', thus ensuring improved learning performance in changing, dynamic environments.[8]

The notion that one may sensibly seek to 'quantify' surprises will raise some philosophers' eyebrows. While there is a clear qualitative difference between the mere 'un-anticipated event (or proposition)' and that which forces us to re-consider our understanding of ourselves and/or the world around us, some will question the wisdom inherent in seeking to quantify the difference between, say, the surprise inherent in hearing oneself sing for the first time and the surprise one may experience in discovering a foreign culture. One may seek to accommodate this incommensurability issue by borrowing Quine's 'sphere of beliefs' metaphor. According to the latter, what matters – and what one might seek to quantify – is how close the unsettled beliefs are to the centre of the sphere. The closer to the centre, the greater the surprise, since it requires one to revise larger chunks of dependent beliefs or propositions. Still, some might argue that it remains unclear where in that sphere profoundly subjective (and aesthetic) experiences are meant to fit: they might have a very significant impact upon one's life, reversing, say, years of systematic self-deprecation (with its concomitant normative stands), yet appear fairly superficial in terms of dependent beliefs.

---

[5] "Being able to detect novel and surprising stimuli is necessary for efficiently learning new memories without altering past useful memories" (Carpenter and Grossberg, 1988).

[6] While there are clear (and interesting) links between the two, one has to distinguish between surprise-related literature that is focused on improving the learning performance of the system (discussed above) and research that is concerned with maintaining the interest of the system's end-users by introducing "serendipitous" outputs, as in recommender systems. For a survey, see (Kotkov, Wang, & Veijalainen, 2016).

[7] (M. Faraji, 2016; M. J. Faraji, Preuschoff, & Gerstner, 2016) "Shannon surprise Eq. (2.1) and Bayesian surprise Eq. (2.3) are two distinct yet complementary approaches for calculating surprise. Shannon surprise is about data as it captures the inherent unexpectedness of a piece of data given a model […] Bayesian surprise is about a model as it measures the change in belief about model parameters." (M. Faraji, 2016, p. 12).

[8] (M. Faraji, 2016, p. 39) "This remaining uncertainty ensures that an organism can still detect a change even after having spent an extensive amount of time in a given environment". [Is the prior citation for this quote as well?]

Could it be that the burgeoning literature on Machine Learning-related 'surprises' (and tools to preserve their possibility) is in fact considering something fundamentally different from that which occupies philosophers when they seek to understand the role played by surprises within normative agency? Section 1.2 considers whether there is an inherent link between a capacity to be surprised and a capacity to form habits.

### 1.2. Habit and Surprise: the flip sides of a coin?

The discussion above proceeded from the assumption that computers are capable of being surprised (hence it makes sense to talk of tools designed to preserve that ability). This section aims to debunk this assumption by highlighting the conceptual link between an ability to be surprised and the nature of the effort needed to overcome habits. Considered independently of that effort, the way in which efficiency concerns typically lead to various optimisation strategies indeed supports the idea that there are such things as `algorithmic habits'.

To make an algorithm run more efficiently, it is indeed standard practice to `profile', i.e. to look at which parts of the software are going fast or slow and store underlying calculations for certain tasks. In a dynamic environment, where the heuristics that are relied on to determine whether such underlying calculations are still valid, over-optimisation will compromise performance, just as over-reliance on habits will compromise human performance. Yet the algorithm-human analogy when it comes to habit only goes so far. The next paragraphs highlight the importance of the qualitative difference in the nature of the mechanisms underlying habit reversal in humans v. algorithms.

In all cases -whether it be repeated movement, posture, calculation or frame of thought, habit requires repetition. In the pattern shaped by this repetition, at some point a habit is formed. Any attempt to identify a precise moment in time when a habit is born is doomed to failure, for diminished awareness of the pattern underlying it is key to its emergence.[9] While one can intentionally seek to develop some habit,[10] the latter is born only once the behaviour, posture or frame of thought underlying it has become so internalized that it takes *effort* to bring it back to *conscious awareness*. This notion of 'effort to bring it back to conscious awareness' is absolutely essential to the existence of a habit: you have not developed a habit until the pattern of behaviour underlying it slips from your consciousness, even if momentarily. Since computers arguably lack 'conscious awareness'[11], one could straight away conclude that they are incapable of forming habits.

But let's – for the sake of argument – adopt an agnostic or lax understanding of consciousness, according to which computers would develop 'conscious' response patterns (based on, say, hyper-personalised anticipation of their users' tasks). The processing *effort* required on the part of computers to overcome such patterns would still be fundamentally different from the kind of effort required of humans to overcome deeply ingrained habits.

---

[9] External observation necessarily comes too late.
[10] Habits can be acquired in many ways: intentionally (for instance to foster the realisation of a particular goal) or unintentionally (through upbringing or simply responding to particular environmental features that shape one's behaviour).
[11] Those who claim that one day, as a result of increased computational power and some rather mysterious 'complexity', computers may well wake up to their own existence  (Du Sautoy, 2016) have yet to specify what distinct 'consciousness enabling' features, if any, such conscious computers would have. (Tallis, 2011) brilliantly exposes the pitfalls (and naivety) of the materialist reductivism that conditions endeavours to 'measure' consciousness or locate its seat in the brain.

This qualitative difference matters, and reflects the role of somatic markers in the process of habit acquisition. According to Damasio's theory (Damasio, 1996) (Damasio, 2008), for each action that is in the process of becoming habitual, the brain accumulates information about the somatic outcomes (what bodily sensations are associated with that action) and encapsulates that information into an intuitive 'marker' that is subsequently activated (and steers behaviour) in any context relevant to that action. Even those who challenge Damasio's somatic marker theory readily concede the essential role played by bodily sensations in the formation -and reversal- of habits. The latter, habit reversal process can be painful (smoking cessation is the easiest example) or utterly disorienting: Proust for instance compares the effect of certain novels to `temporary bereavements, abolishing habit'.[12] Might digital computers ever experience the process necessary to habit reversal in such a `bereavement-like' fashion?[13] It would be pure speculation -and thus somewhat pointless- to try to answer that question today. The amplitude of the challenge at stake however warrants scepticism when it comes to the purported ability of digital computers to develop habits –and hence to be surprised in a non-trivial sense.

Now, some might want to ask: why should this inability to develop habits matter at all in terms of one's capacity to be surprised? One might think that this inability is a good thing, given the extent to which habits can stand in the way of surprises. Yet habits do not only compromise our ability to be surprised. They also enable it. Without habits, we would be perpetually clueless. Habits not only shape and determine our sense of self, they are also at the root of our understanding of the world, both as it is and as it should be. To concede that habits are at the root of most of our normative stands, determining what values we endorse and what type of life is seen as worth living, goes against the dominant, intellectualist tradition. The latter likes to think that it's our conscious, deliberative self that is exclusively in charge – at least when it comes to ethics and morality. We have known for some time that this is not the case. Yet even now that we have extensive evidence suggesting that culturally acquired habits of evaluation and the intuitions they give rise to have a direct, causal impact on most of our moral judgments (Haidt, 2008), the impact of this intellectualist tradition remains considerable. Studies of habit within legal and moral philosophy indeed remain few and far between, and have yet to be taken on board by those computer scientists delving into the need for 'value-aligned systems'.

---

[12] (Proust, 1996, p. 642)

[13] In the context of a workshop discussion, Chris Baber suggested the possibility of designing a computer system in such a way as for the latter to trigger some electrical short-circuit whenever a habit (or repeated pattern) needs to be reversed: this is one possible way in which one may seek to mirror the human experience of habit reversal – it is of course an open question whether such `mirroring' ambition is desirable (see section 2.2. for possible arguments backing such mirroring aim).

If the ability to be surprised is indeed concomitant with a capacity to develop habits, could it be that the blossoming Machine-Learning literature which emphasises the role of (and need to preserve) the ability to be 'surprised' is in fact talking about something else? As a distinct but closely related concept, 'artificial curiosity' (Ngo, Luciw, Foerster, & Schmidhuber, 2012; Storck, Hochreiter, & Schmidhuber, 1995) better captures the concerns of those seeking to improve the learning performance of Machine learning applications. Because it is designed to rely on the information that is gathered at each stage to adapt or fine-tune its parameters, a learning algorithm is by necessity only as good as the data it has been fed. If it has to function in a dynamic environment, or at least one that is not as stable as anticipated, the learning algorithm is likely to produce sub-optimal results. The problem is that the algorithm itself is unlikely to be able to grasp the instability of the environment within which it is made to function, unless extra, 'artificial curiosity' constraints are introduced. The latter would prompt the system to actively extend the range of instances over which they have data and look for uncommon, 'black swan events'[14] that might demand some model alteration (thus counter-balancing the Bayesian tendency towards near-zero model uncertainty).

The resulting dynamism of the process underlying model formation (and alteration) is important in more than one respect. Aside from improving the performance of the system itself, I argue that such dynamism conditions the very possibility of ever deploying autonomous systems meant for morally loaded contexts, given that ethics cannot but remain a work in progress. This argument -'Thesis 2'- is unpacked in the following section.

---

[14] The data-mining literature that focuses on the detection of anomalies within noisy data-sets (Eskin, 2000) proceeds from a different starting point (in that the data is given) but the underlying logic is similar.

## 2. Autonomous systems fit for dynamic, morally-loaded contexts

This section analyses the implications of the following two theses taken together:

*Thesis 1*: Without the depth of habit to emotionally (and somatically) 'anchor' model certainty, a computer's experience of something new cannot but remain very different from that which in humans gives rise to non-trivial surprises. This thesis was the focus of section 1.

*Thesis 2*: Ethics cannot but remain a work in progress. This thesis (henceforth the 'work in progress view') reflects an understanding of moral values that is often left unarticulated, especially by those who support the opposite view, according to which a final, definitive answer to the 'how should we [I] live' question is both desirable and in principle available to us (henceforth the 'final view'). The latter stand silently underlies much of the literature on the so-called 'value-alignment problem' (led by the work of Bostrom[15] and others).

Section 2.1. unpacks the implications of either account of moral values ('final' v. 'work in progress') when it comes to discussing the possibility of -and design challenges inherent in- autonomous systems meant for morally loaded contexts. Section 2.2. moves on to consider the desirability of such systems if theses 1 and 2 are both correct.

### 2.1. Beyond axiological identification and incorporation: the need for moral change mechanisms

Current efforts to ethically 'train' (or constrain) autonomous systems that are capable of being deployed in morally loaded contexts pay little attention -if any- to the difficulties that stem from the unavoidable need for moral change. The challenge inherent in those systems having to take into account a wide range of moral values is often referred to as the 'value-

---

[15] (Bostrom, 2014)

alignment problem'. Its discussion is currently dominated by the assumption that one may validly approach this problem as one involving the two first challenges below:

-*Challenge 1: Identification*. One needs to somehow identify a set of 'valid' moral values that are to act as constraints or guiding principles in the operation of the system.

-*Challenge 2: Incorporation*. In contrast to 'top-down' incorporation strategies[16], those systems which proceed on the basis of reinforcement learning strategies (effectively merging the identification and incorporation challenges) have the merit of being compatible with an acknowledgment of the dynamic nature of moral values, but fail to take on board the asymmetry in the nature of the mechanisms that may plausibly underlie moral change in humans v. computer systems. This asymmetry is at the root of the unacknowledged, third challenge -'Mechanisms for change'- discussed below.

Before turning to the latter, it is important to emphasise that the task of identifying and then incorporating a set of moral values that operates as relevant constraints on an autonomous system is not particularly affected by the extent to which one upholds the 'final' or 'work in progress' understanding of moral values. Whether or not one entertains the idea that a 'final', uniquely correct answer to the question of how we should live together is both desirable and in principle available, as a matter of fact that answer tends to be the object of controversy and disagreement in any society. Researchers working on the so-called 'value-alignment problem' (or 'value-loading problem'[17]) acknowledge this, and there is already significant work endeavouring to tackle the issues raised by the contested nature of relevant moral values.[18] Along this line, reinforcement learning (RL) strategies may be deemed to stem at least in part from an effort to circumvent the vexed 'identification challenge': instead of coding ethical values 'by hand', the idea is to let data train the system.

---

[16] This focus on top-down incorporation strategies is openly visible in the 2016 IEEE report: `The conceptual complexities surrounding what "values" are make it currently difficult to envision AIS that have computational structures directly corresponding to values. However, it is a realistic goal to embed explicit norms into such systems, because norms can be considered instructions to act in defined ways in defined contexts.' (IEEE, 2016, p. 22). See also (Anderson and Anderson 2011) (Arnold, Kasenberg, & Scheutz, 2017) for the perspective of `adding' or incorporating `top-down' some kind of ethics to a system's decision-making procedures.
[17] (Bostrom, 2014)
[18] For a discussion of the concrete challenges raised by the "contested" nature of the moral values informing algorithmic content moderation, see (Binns, Veale, Van Kleek, & Shadboldt, 2017).

Yet in contrast to both supervised and unsupervised learning approaches, the set of data on the basis of which RL methods proceed is not given a priori: the data is generated by the artificial agent's interaction with the environment. The aim of the learning process is to come up with an action-selection policy that minimises some measure of long-term cost, which is determined on the basis of a -continuously updated- utility function. Aside from the difficulty inherent in articulating the initial utility function, traditional reinforcement learning methods are vulnerable to deception on the part of the system: 'an AI system might manipulate its reward functions in order to accomplish the goals that it holds as most important, however unethical its effects on human beings'.[19]

Inverse reinforcement learning methods, by contrast, do not proceed from a given, initial utility function: the system is meant to infer the latter function from observed behaviour. Russell et al.[20] propose this behaviourist, bottom-up approach as a way of approximating our expectations for an ethical system: as these inferred expectations evolve, the system is meant to update its utility function accordingly, thus in principle solving the 'mechanisms for change' challenge discussed below. Yet there are two fairly major difficulties inherent in this approach: first, one cannot but dangerously over-simplify (and distort) ethical aspirations if one allows observed behaviour to be their sole determinant. The IRL method is also likely to heavily reinforce the status-quo: this is in part because it is unlikely to pick up the significance of seemingly isolated civil disobedience or morally courageous acts. It is also because (just like other methods), if successful, its freeing us from the normative work required to answer the 'how should we [I] live?' question may leave us content to 'tag along', unable to appreciate the very point of engaging with such a question.

-Challenge 3: Mechanisms for change.

For those who entertain the idea that a final answer to the 'How should I [we] live?' question is not only available in principle, but desirable, the prospect of developing some superintelligence on whose superior cognitive capacities we could rely on 'to figure out just which actions fit [what is morally right]'[21] is extremely attractive:

> *'The idea is that we humans have an imperfect understanding of what is right and wrong, and perhaps an even poorer understanding of how the concept of moral rightness is to be philosophically analyzed: but a superintelligence could understand these things better'.*[22]

---

[19] (Arnold et al., 2017)
[20] (Russell, Dewey, & Tegmark, 2015)

The notion that there is a concept of 'moral rightness' whose contours do not depend in any way on our all too human, fallible, short-sighted nature has a long pedigree in the history of moral philosophy (its roots can be found in Plato). On this account, all we need to rescue us from our persistent moral failings is a once-and-for-all source of enlightenment. A superintelligence that does not share in any of our shortcomings -biological or otherwise- could provide precisely that, and more (it might also figure out a way to motivate us to act according to 'moral rightness'). There is neither room nor need, on this account, for any 'mechanism for moral change': life's circumstances might change, but 'moral rightness' does not…Or does it?

The tradition that questions the extent to which one may meaningfully speak of 'moral rightness' independently of the kind of creatures we are (itself a work in progress) is almost as old as Plato -its roots can be traced back to Aristotle's moral psychology. The tricky part, on this account, is to avoid throwing the baby out with the bathwater: there is a crucial difference between asserting the dependency of moral rightness upon who we are and a relativist 'anything goes'. Many have fallen for the mistake of assuming that without moral realism there is no ethical objectivity to be had.[23] Bostrom has the merit of explicitly articulating this assumption when he states:

---

[21] (Bostrom, 2014)
[22] (Bostrom, 2014)
[23] Mackie's `error theory' has been influential in legal (and moral) theory (Mackie, 1990) and is indirectly referred to by (Bostrom, 2014). (Putnam, 2004) exposes the extent to which this assumption reflects a Cartesian dualism according to which there is only one sort of objectivity, that of the natural sciences.

> *'What if we are not sure whether moral realism is true? We could still attempt the [Moral Rightness] proposal […] we could stipulate that if the AI estimates with a sufficient probability that there are no suitable non-relative truths about moral rightness, then it should revert to implementing coherent extrapolated volition[24] instead, or simply shut itself down'.[25]*

Suppose the AI were indeed to shut itself down. What would it leave us with? We'd still be trying to find our way around the world, generally aiming for better (rather than worse) ways of living together. Rather than dismiss as 'lacking in objectivity' the rich background of norms that informs our ongoing ethical efforts, a (non-reductive) naturalism starts from precisely such 'contingent' normative practices. Eminently fallible, our answers to the 'How should I [we] live?' question cannot but be constantly changing, just as human nature evolves as we learn to live together. In such a dynamic normative context, the process through which we engage with the ethical question matters at least as much as the answer itself, for that process renews the background of normative practices that informs others' ethical judgment.

Now imagine that, at some point in the future, a superintelligence is somehow developed along the lines considered by Bostrom. Whether it is supposed to have 'cracked' moral rightness for us or relies instead on our 'extrapolated' wish -'if we were more the people we wished we were'[26], that superintelligence won't fool around. It will set us on the path to 'righteousness', like it or not. Even if it were to find a palatable way of imposing what may otherwise appear to us as morally alien and abhorrent (hence avoiding potential 'human override' procedures), in its bid to achieve moral perfection it may well end up depriving us from the possibility of living ethical lives. The systematic offloading of the normative work[27] required to answer the 'how should we [I] live?' question to an AI may indeed leave us incapable of appreciating the very point of engaging with such a question. Thus the cost of moral perfectionism (and AI-enabled normative laziness) may turn out to be the end of ethics: we might be normative animals, but without regular exercise, our moral muscles will just wither away, leaving us unable to consider alternative, better ways of living together.

---

[24] (Yudkowsky, 2004) defines our `coherent extrapolated volition' (which a superintelligence would be relied on to figure out, and implement) thus: `Our coherent extrapolated volition is our wish if we knew more, thought faster, were more the people we wished we were, had grown up farther together; where the extrapolation converges rather than diverges, where our wishes cohere rather than interfere; extrapolated as we wish that extrapolated, interpreted as we wish that interpreted'.

[25] (Bostrom, 2014, p. loc. 5058)

[26] (Yudkowsky, 2004)

[27] Bostrom's characterisation of this work in purely cognitive terms reflects his robust realist meta-ethical premises.

Of course we need not adhere to the 'final' understanding of moral values referred to earlier. What would an autonomous system meant for morally loaded contexts look like, if we start from the opposite, 'ethics as a work in progress' conception? Such a system would have to start from somewhere. Even if -and this is an ideal, 'sci-fi' scenario at the moment- that system were to learn to value things sufficiently slowly and progressively as to mimic the human process of growing up, it is unclear whether such a system could be said to be capable of developing habits (see section 1).[28]

Without habits to somatically anchor deeply held moral stances, the processes through which an automated system might be led to change and adapt its moral outlook is bound to differ fundamentally from that of humans'. The comparative ease (and speed) with which such systems would adapt to novel, ethically challenging situations may turn out to be a mixed blessing. It will, in any case, confront us, humans, with creatures that may originally have been trained to think and feel like us (and this entails some ostensible fallibility), but which will potentially develop into creatures holding seemingly 'alien' views…Just like teenagers? Rini draws on this analogy with teenagers to argue that we ought to accept the fact that our 'artificial progeny might make moral choices that look strange'.[29] In the next section I argue that we'd be lucky if this relatively benign analogy holds true.

To sum up: one cannot discuss the possibility of developing autonomous systems meant for morally loaded contexts without taking on board the fact that we humans do keep changing our moral stances. For those who believe in the possibility and desirability of a 'final' answer to ethics, that fact merely reflects our all too fallible and fickle nature. From that perspective, the prospect of somehow being able to rely on a system's superior cognitive prowess to figure it all out for us, once and for all, is a boon that ought to be met with enthusiasm. From the 'ethics as a work in progress' perspective, by contrast, such a prospect can only be met with scepticism at best (or alarm at worst).

---

[28] This would depend on the kind of effort required on the part of such system to shake off any `habituated' pattern of behaviour (or thought). If there is a qualitative difference between the effort required to overcome such patterns and the processing effort concomitant with just any other task, then we might ask ourselves if that system has indeed developed a habit. For the reasons outlined in Section 1, I remain sceptical about the extent to which automated systems may do so.

[29] (Rini, 2017)

In terms of feasibility, the challenge inherent in artificially introducing 'mechanisms for change' that would enable a system to continuously update its moral stances in a manner that is both accurate and appropriate cannot be overestimated. Why? Either one seeks to build a system that somehow reflects the way in which we humans are constantly in the process of addressing the 'how should I [we] live?' question. Given the considerable tensions and discrepancies inherent in that process, that reflection would be a very rough approximation at best – or a dangerously reductivist simplification at worst (see the concerns related to the inverse reinforcement learning method, above). Or one allows the system to 'leap ahead' of us and become an advocate for the need for change when we haven't quite caught up with its necessity. If the (benign) analogy to teenagers is to hold true, however, they can only leap so far before we lose the ability to have intelligible conversations.

Whether it is met with enthusiasm or scepticism, the ambition to construct autonomous systems meant for morally loaded contexts comes with a hazard that is seldom considered: put lazy normative animals -that's us- together with systems to which we may offload the task of figuring out the 'how should I [we] live?' question, and what you get are endless moral holidays[30], and lazy animals *tout court*. Our capacity for normative reflection - querying how the world could be made better, rather than 'sitting on it'[31] - is all too often taken for granted. What if that capacity can be lost through lack of normative exercise? What if we enjoy the comforts of automated, simplified practical reasoning a bit too much, a bit too long? What was meant to be a 'moral holiday' may turn out to be a condition which we are unable to get out of, for want of being able to mobilise moral muscles that have become atrophied through lack of exercise.

---

[30] This concept of `moral holidays' is borrowed from (James, 2000). The following passage highlights its relationship to what James calls `absolutism', or what I refer to as a `final' understanding of ethics: `[The world of pluralism] is always vulnerable, for some part may go astray; and having no 'eternal' edition of it to draw comfort from, its partisans must always feel to some degree insecure. If, as pluralists, we grant ourselves moral holidays, they can only be provisional breathing-spells, intended to refresh us for the morrow's fight. This forms one permanent inferiority of pluralism from the pragmatic point of view. It has no saving message for incurably sick souls. Absolutism, among its other messages, has that message […] That constitutes its chief superiority and is the source of its religious power. That is why, desiring to do it full justice, I valued its aptitude for moral-holiday giving so highly' (James, 1911).

The prospect of AI-enabled, extensive moral holidays may sound like too exotic a possibility to worry about its effects on our capacity for normative agency. Indeed there are reasons to doubt the feasibility of developing such autonomous moral 'agents', even in the longer term.[32] Yet we do not need to settle this feasibility issue to draw a normative conclusion that has an impact upon the development of systems that are already in the process of changing the way we make (sometimes morally-loaded) decisions: the whole of section 3 is devoted to the much more 'down to earth' task of designing decision-support systems that support and foster our normative agency, rather than compromise it. Before that, section 2.2. questions the desirability of fully autonomous moral agents.

### 2.2. No Pinocchio, no teenagers either: questioning the desirability of fully autonomous artificial moral agents

Who wouldn't be fascinated by the prospect of being able to engineer creatures designed to overcome human frailties and limitations -including, most importantly, our very limited ability to store and process data? It is not difficult to explain our captivation for the debate surrounding the possibility of developing fully autonomous artificial agents that are capable of acting and thinking 'like us' (a debate which Turing[33] launched almost 70 years ago). Because many identify our normative inclinations as a peculiarly human trait -most of us are not content with the world 'as it is', and keep wondering how it could be made better, normative agency has become a sort of 'yardstick'. If we can develop artificial agents that are capable of thinking 'normatively', then we'll have cracked the challenge set by Turing[34] in the 1950's: we'll have created creatures that are truly 'like us'. Should we aim to do so?

---

[31] When thinking of atrophied moral muscles etc., the image I associate with this comes from the film `Wall-E', depicting ballooned humans each sipping their smoothie while watching a movie on some floating cushion: due to lack of exercise, they are simply unable to stand up and have become utterly dependent on some automated entertainment structure.

[32] (W. Wallach & Allen, 2013) `accept that full-blown moral agency (which depends on strong A.I.) or even "weak" A.I. that is nevertheless powerful enough to pass the Turing test...may be beyond current or even future technology'.

[33] (Turing, 1950)

Aside from fulfilling some dubious 'godlike' fantasy, it is far from clear why we would want artificial, autonomous agents to be 'like us'. Indeed, when it comes ethical concerns, we are likely to be better off if such systems are significantly different from us. Different how? If one adheres to the idea that there is a 'final' answer to the ethical question, the extent to which they'll end up different from us depends on how far we -fallible human beings- have strayed from the path of 'righteousness'. If we've strayed far, we might not like what we get…

The latter conclusion is not very different if we start from an 'ethics as a work in progress' understanding instead. Seen from that perspective, the challenge consists less in determining what moral stands autonomous systems should start with, and more in figuring out how such systems are to evolve in a way that is compatible with our constantly evolving attempts at coming to grips with the 'how should I [we] live?' question. Given the very different way in which such systems experience novelty (see section 1), they are sure to evolve differently from us, no matter how much we might try to shape them in our image. For Rini, we should embrace this prospect of likely divergence:

> *'If we are wise and benevolent, we will have prepared the way for them to make their own choices – just as we do with our adolescent children. What does this mean in practice? It means being ready to accept that machines might eventually make moral decisions that none of us find acceptable. The only condition is that they must be able to give intelligible reasons for what they're doing'.*[35]

---

[34] To bring home the roots of the debate about artificial moral agency, some speak of the `moral Turing Test': *'A moral Turing test (MTT) might similarly be proposed to bypass disagreements about ethical standards by restricting the standard Turing test to conversations about morality. If human interrogators cannot identify the machine at above chance accuracy, then the machine is, on this criterion, a moral agent'* (Allen, Varner, & Zinser, 2000). Allen et al. acknowledge that there are some obstacles inherent in this test being used a benchmark for the development of artificial moral agents: aside from the fact that the moral standards relevant to such agents may have to be more demanding than those that apply to us, moral agency can't all be a matter of the reasons one gives (but one could tweak the test to allow for comparisons of *actions* between human and artificial moral agent). (Scheutz & Arnold, 2016) go further, and highlight, among the problems inherent the MTT, the fact that the MTT `if it carries enough similarity to the original Turing test to deserve that name, ultimately and unavoidably rests on *imitation* as a criterion for moral performance. In turn, the kind of deceptive responses consistent with imitation, both as a representation of the agent and as a substitute for moral action writ large, undermines a more accountable, systematic design approach to autonomous systems'. I would go further still, and emphasise what a bizarre understanding of moral agency the MTT conveys: unlike thinking (which was the focus of the original Turing test), moral agency is not a predicate for which we lack some essential criteria. All humans think (bar marginal cases). But do all humans exercise their moral agency? No. It is certainly a peculiarly human trait that we are capable of moral agency. When we do exercise that capability, we deploy it in myriad different ways. Can we lose that capability? Yes. In fact, in the `endless moral holidays' scenario I have described earlier, one could envisage a reversed `moral Turing Test', whereby a computer is asked to `blindly' interrogate a human and a computer in a bid to determine which is which: they might find that test much easier than humans do…

While I would embrace such a liberal stand when it comes to teenagers, for artificial autonomous agents it's a punt too far, as far as I am concerned. Difference can be good. Indeed, confrontation may be salutary: in the next section I highlight the benefits of designing decision-support systems that are capable of shaking us out of our moral torpor (this might involve some sort of confrontation). But we don't need fully autonomous moral agents for that. All we need is a willingness to meet 'others': fellow human beings.

## 3. Apt at taking us by surprise? Autonomous v. decision support-systems

> *'[R]obots will have particular jobs before being general "actors" in society. So instead of seeking a universal sense of "morality" across all possible contexts, should moral competence not be confined to the particular role (and imitation) that the robots is designed to fill? […] Robots will not likely emerge from manufacture as free-ranging citizens of the world, with no particular vocation or role to define their actions and decision-making. While the range of social robots can be wide, one must ask what limited set of tasks or objectives they are at minimum designed to accomplish.'*[36]

---

[35] (Rini, 2017)
[36] (Scheutz & Arnold, 2016, p. 28)

## 3.1. Decision-support systems: Surprising by design?

Just as the early days of AI research spurred on a fruitful renewal of the (old) philosophical debate about the nature of expertise[37], I believe that recent attempts at tackling what computer scientists call the 'value-alignment problem' are similarly lending renewed vigor to philosophical reflection about the distinctiveness of ethical expertise. The latter was first (and most famously) highlighted by Socrates: while one may inquire into virtue as if it were, or at least resembles, the 'expert knowledge of living well', Socrates reminds us that virtue cannot be *taught* in the same way as carpentry or seafaring (or other *techne*) might. If virtue involves any form of knowledge, at its core is the correct appraisal of what one does not know.[38]

This openness to being 'called into question' (and possibly surprised) is similarly deemed to be central to more recent accounts of ethical expertise:

> *'Ethical expertise, in the everyday sense, is possible only through a continual openness to an experience of self-doubt reminiscent of Levinas' idea that we are 'called into question' by the other, and therefore very different from what is involved in learning to become expert in other skills' (Reed, 2013, p. 247)*

---

[37] (Dreyfus, 1965, 2005)
[38] Socrates denies that he possesses any expertise in *arete*, often translated as "virtue". Many interpreters question the sincerity of Socrates' professed ignorance. Yet one may take this professed ignorance to be merely disclaiming certain or "expert" knowledge (as in "*techne*") while acknowledging "nonexpert" or "human" wisdom: *"What kind of wisdom? Human wisdom, perhaps. It may be that I really possess this, while those whom I mentioned just now are wise with a wisdom more than human; else I cannot explain it, for I certainly do not possess it, and whoever says I do is lying and trying to slander me." (Apology, 20d7-e3)*
The "it" which Socrates otherwise cannot explain is the Delphic oracle's response, according to which no man was wiser than Socrates. Socrates is puzzled by the oracle's claim because "I realize that I am wise concerning nothing great or small" (*Apology*, 21b4-5). Socrates resolves his puzzlement thus: what makes no one wiser than him is his correct appraisal of what he does not know. *"What makes him the wisest of the Greeks –what he shares with no one he has yet met- is his recognition that he fails to know anything fine and good. Some of those he has met know things he does not know. But all of them think they know things fine and good, the most important things, when they do not. Socrates, alone of the Greeks, fails to have this false belief […] He alone realizes that `in truth he is worthless with respect to wisdom'"*.

Yet so far the debate about designing decision-support systems in value-loaded applications has paid little attention to the extent to which such systems may be designed to foster this openness to being called into question. The current focus on various value-incorporation strategies in fact pulls very much in the other direction. Whenever such strategies succeed in enabling us to step back and relax - somehow trusting machines to have gotten our 'moral sums' right-, they cannot but compromise the kind of critical engagement that is essential to retaining an ability to being called into question. To illustrate the extent to which the latter ability is all too easily compromised (even in an 'off-line, low-tech' environment), consider the following quote:

> *' [T]he horrible thing about all legal officials, even the best, about all judges, magistrates, barristers, detectives, and policemen, is not that they are wicked (some of them are good), not that they are stupid (several of them are quite intelligent). It is simply that they have got used to it. Strictly they do not see the prisoner in the dock; all they see is the usual man in the usual place. They do not see the awful court of judgment; they only see their own workshop.' (Chesterton, 1955)*

What the above-mentioned 'legal officials' lack is not cognitive prowess but rather a willingness to let the words and presence of others reach them. Highly habituated, frequently activated cognitive skills are both efficient and comfortable. They are also very good at warding off as irrelevant factors that might otherwise call on our ethical responsibility - and perhaps demand moral change. When the emotional de-sensitization concomitant with professional habituation is combined with the effects of a normative structure that is designed to simplify our practical reasoning (a legal system fits the bill, but decision-support systems do too), we are likely to get the extensive 'moral holidays' scenario described above.

The answer is not to ditch any form decision-support, but rather to design the latter differently. Of particular interest, in terms of method, are systems that place (and retain) end-users within the learning loop (this approach is sometimes referred to as 'interactive machine learning' or 'IML'[39]). An explicit requirement to keep monitoring the result of the learning process, combined with a demand for regular input on the part of end-users, has the potential to not only improve the system's learning performance; it might also keep moral torpor at bay: For IML to develop into a -however partial- answer to the 'moral muscle atrophy' problem discussed in 2.1, one may for instance consider the introduction of emotionally charged, provocative scenarios that are meant to get end-users to consider critically the value choices informing the algorithms they routinely rely on.

In the context of applications designed to support professionals (medics, lawyers etc.), short video clips that are meant to induce some perspective reversal (with or without immersive Virtual Reality[40] tools) could prove very effective. Ethical lapses within professional practice indeed most often stem from a failure to discern ethically relevant considerations which may only be distantly connected to the problem in relation to which a professional is consulted. Whether it comes to the need to take into account the vulnerability of a patient's family member[41], say, or considering the impact of a company's merger upon the environment and members of the local community, an ability to see beyond one's immediate query does condition the ethical awareness which a professional needs if she is to live up to her particular responsibility.[42]

Professions-specific automated systems can and should be designed with a view to fostering such perspective widening.[43] They may also usefully leverage recent research on the factors that impact upon individuals' differential creativity. Among the characteristics used to assess such creativity, fluency and flexibility are of particular relevance when it comes to counter the effects of professional routine. Similarly, they can and should encourage an 'ethical feedback loop' that carves a continuous, active role on the part of whichever professional community an automated system is designed for. The latter feedback may allow for some dynamic process of adaptation to the changing values of end-users, thus addressing the issue related to the dynamics of moral change addressed in section 2.1. Yet aside from some research focusing on the challenges raised by multiple people interacting with machine learning systems[44], there does not seem to be much published in this 'interactive machine learning' area recently, which is a shame.

---

[39] "Although humans are an integral part of the learning process (the provide labels, rankings etc.), traditional machine learning systems used in these applications are agnostic to the fact that inputs/outputs are from/for humans. In contrast, interactive machine learning places end-users in the learning loop (end users is an integral part of the learning process), observing the result of learning and providing input meant to improve the learning outcome. Canonical applications of IML include scenarios involving humans interacting with robots to teach them to perform certain tasks, humans helping virtual agents play computer games by giving them feedback on their performance." (Wallach, Wendell, & Allen, 2008)

[40] To foster such perspective reversal (initially in the context of psycho-therapy), Mel Slater has pioneered the use of embodiment techniques (relying on sophisticated immersive virtual reality tools). See (Falconer et al., 2014; Osimo, Pizarro, Spanlang, & Slater, 2015)

[41] For a study examining the impact of expertise and cognitive load upon a GP's ability to pick up signs of child-safeguarding concerns, see (Pan et al., 2018)

[42] The nature (and ethical grounds) of professional responsibility are discussed at length in [...]

[43] They may also usefully leverage recent research on the factors that impact upon individuals' differential creativity (Zabelina, L. Robinson, D. Council, R, & Bresin, 2011). Among the characteristics used to assess such , fluency and flexibility are of particular relevance when it comes to counter the effects of professional routinisation.

### 3.2. Human encounters with 'relatively autonomous' systems: a source of ethically relevant surprises?

Those who doubt that a computer may ever be able to surprise us tend to be influenced by the fact that computers are, at bottom, 'made by us'. The idea that only something that comes from 'outside us' may surprise us is related to the romantic intuition that underlies Lady Lovelace's objection. According to the latter, true autonomy requires the ability to originate something new in a radically unprecedented way. To the extent that the intelligence displayed by computers is not only *enabled*, but also -within 'traditional' algorithms at least- *manufactured* by us, non-trivial surprises should not arise, or so the reasoning goes. That reasoning is flawed in two respects. First, it proceeds from the naïve assumption[45] that one may indeed delineate a sphere of reality that is wholly 'outside us', uncontaminated by the human touch (and thus apt to surprise us). Equally problematic is the assumption that our status as manufacturers entails that we cannot but have a perfect grasp of both the algorithm's cognitive ramifications and its potential deployment and impact upon our world.

---

[44] `An important opportunity exists to investigate how crowds of people might collaboratively drive interactive machine learning systems, potentially scaling up the impact of such systems. […] in understanding how we can coordinate the efforts of multiple people interacting with machine learning systems.' (Amershi, Camak, Knox, & Kulesza, 2014)
[45] That assumption itself proceeds from a Cartesian dualism that aims to neatly distinguish the `world in and of itself' from human impressions.

Now, the idea that non-trivial surprises cannot arise from computers that are 'made by us' not only proceeds from flawed assumptions. Today it is also fundamentally outdated. It may only have become fashionable fairly recently, but Machine Learning as a broad methodology has been around for some time. According to Mitchell, what distinguishes that method from 'traditional' computer science (which 'focuse[s] primarily on how to manually program computers') is the 'focu[s] on the question of how to get computers to program themselves' (Tom Michael Mitchell, 2006). This way of characterising Machine Learning puts a lot of emphasis on the relative autonomy of the system - perhaps overly so: humans indeed play a determinant role in 'designing' both the system and -to a varying extent- the data it learns from, particularly in so-called supervised learning methods.[46] The importance of human input is better captured in Mitchell's much-quoted account, according to which 'a computer program is said to learn from experience E with respect to some class of tasks T and performance measure P, if its performance at tasks T, as measured by P, improves with experience E' (T.M. Mitchell, 1997). Both the tasks and, crucially, the performance measure are determined by humans. Even the 'experience' itself tends to necessitate some form of processing by humans: raw data is indeed messy and often needs to be packaged in order for the system to pick relevant or 'useful' variables.[47]

Despite all that, Machine Learning can be said to introduce a degree of *relative* autonomy which does contribute to deflating the appeal of the misguided (and romantic) notion that autonomy can only be considered as a radical, either/or proposition: either one is totally free of prior norms or constraints, and hence autonomous, or one is not.[48] In its current

---

[46] Handwriting recognition is one of the tasks that lends itself to a *supervised learning* approach: one might feed a system with a set of example pairs (x, y) where "x" corresponds to the images containing handwriting and "y" identifies which character is being read. The aim of the learning process is to find a function f:X →Y that matches the example pairs. In *unsupervised learning*, by contrast, some unlabelled data x is given, and the aim is to infer a function that reveals some hidden structure within the data. In *reinforcement learning*, the set of data x is not given. Instead it is generated by an agent's interaction with the environment. The aim of the learning process is to come up with an action-selection policy that minimises some measure of long-term cost.
[47] For further discussion see (Veale, 2017)
[48] `What Lovelace is looking for may require a kind of autonomy that is beyond the bounds of ordinary causation and mathematics. The notion that creativity requires autonomy is one anticipated, at least in nascent terms, by Hofstadter (1995, 411), who seems confident that computation will ultimately be up to the task of capturing the kind of autonomy creative humans exploit. But what if the kind of `thinking for oneself' required by LT entails a form of autonomy known as agent causation? The doctrine of agent causation, which is set out, defended, and shown to be beyond ordinary computation in `Chapter VIII: Free Will' of Bringsjord 1992, entails the view that persons bring about certain states of affairs (e.g. mental events like decisions) directly, with no ordinary physical causal chain in the picture.' (Bringsjord, Bello, & Ferrucci, 2001, p. 25)

form, Machine Learning aims to approximate the way in which we humans typically do update the models governing our understanding of what we do (and who we are) in the light of the data we are confronted with. Because and to the extent that they succeed in this approximation, machine learning applications have the potential to teach us things we never anticipated (or did not know we wanted to know), both about ourselves, and about the way we run our lives.

In the world of games, Machine Learning's recent forays into the world of GO has already contributed novel ways of approaching (and playing) the game (Silver et al., 2016). Might something similar happen to the way we navigate our way around -and structure normative claims about- our world? Could our growing interaction with machine learning applications, which may be said to present us with a very sharp (or accelerated) mirror, generate unexpected insights into who we are, and what we care about? This kind of question has been asked before, but until recently it was mostly confined to the realms of 'what happens if we encounter aliens', science-fiction accounts. To confront instead this question in the context of our encounters with machines that we are very much in the process of designing is both exciting and daunting, given the amplitude of the responsibility it entails.

Design those machines so as to maximise the extent to which we may offload thorny ethical issues, and we may grant ourselves 'moral holidays' - at a cost. Design them instead to periodically jolt us out of our moral torpor, and we might retain enough moral muscle to be able to stand up and question our social practices when they are wanting. Decision-support systems can and should be designed to that effect. What about fully autonomous, artificial agents? If ethical agency is defined by reference to our ongoing endeavour to answer the 'how should I/we live?' question, it is far from clear whether such autonomous artificial agents (if indeed they were to see light of day) could accurately be described as partaking in that ongoing endeavour. No matter how much we try to shape them in our image, their experiencing novelty in a way that is so different from us makes such autonomous artificial agents likely to evolve past the point of mutual intelligibility. Without the latter, it is shock, rather than surprise, that awaits us.

## Conclusion

There is much to learn from what Turing too hastily dismissed as Lady Lovelace's 'objection': 'digital computers' can indeed surprise us. And that's not merely because our calculations or predictions fail more often than computers' do. Prediction failures are expected. Just like a piece of art, a film or a novel, computers (or their algorithms) can be designed in such a way as to lead us to question our understanding of the world, or our place within it. Some humans do lose the capacity to be surprised in that way: it might be fear, or it might be the comfort of all-encompassing, ideological certainties. As lazy normative animals, we do need to be able to rely on various authorities to simplify our practical reasoning: that's ok. Yet the growing sophistication of computer systems designed to systematically free us from the constraints of normative engagement may well take us past a point of no-return: what if, through lack of normative exercise, our 'moral muscles' became so atrophied as to leave us unable to stand against our social practices when they are wanting?

Some may welcome the above scenario: if our normative laziness stems from the comforting knowledge that we are safely in the care of an all-knowing, benign 'superintelligence' setting us on the path to moral righteousness, why worry? In contrast to those who believe that a final answer to the 'how should we live?' question is both available in principle and desirable, others insist that ethics -just like human nature- cannot but remain a work-in-progress. So they worry: the cost of some AI-enabled moral perfectionism might well be the end of ethics. It is uncommon for the implications of contrasting meta-ethical stands to find such concrete illustration. This paper has not only tried to outline those implications within the choices underlying the design of both (putative) autonomous artificial agents and decision-support systems meant for morally loaded contexts. It also makes two distinct normative claims:

1. Decision-support systems should be designed with a view to regularly jolting us out of our moral torpor. In this context, the potential of various interactive machine learning methodologies may prove a fruitful avenue for further research.

2. Without the depth of habit to emotionally (and somatically) 'anchor' model certainty, a computer's experience of something new cannot but remain very different from that which in humans gives rise to non-trivial surprises. This asymmetry has important repercussions when it comes to ethical agency, and the shape the latter would take in so-called 'artificial moral agents': it is not just that they would be likely to 'leap morally ahead' of us, unencumbered by the weight of habits (if it were, mutual intelligibility would merely be a matter of decelerating evolution). The main reason to doubt that the moral trajectories of humans v. autonomous systems might remain compatible stems from the fundamental asymmetry in the mechanisms underlying moral change. Whereas for humans surprises and emotional encounters will continue to play an important role in waking us to the need for moral change (for as long as we remain capable of ethics), cognitive processes are likely to rule when it comes to machines. This asymmetry cannot but translate into increasingly different moral outlooks, to the point of (likely) unintelligibility. The latter prospect is enough to doubt the desirability of autonomous moral agents.

IEEE. (2016). *Ethically Aligned Design: A vision for prioritizing human wellbeing with artificial intelligence and autonomous systems (version 1)*. Retrieved from <http://standards.ieee.org/develop/indconn/ec/ead_v1.pdf>

James, W. (1911). "The Absolute and the Strenuous Life" *The Meaning of Truth* (pp. 226-229). New York: Longman Green and Co.

James, W. (2000). Pragmatism: a new name for some old ways of thinking *Pragmatism and other writings*: Penguin Classics.

Kotkov, D., Wang, S., & Veijalainen, J. (2016). A survey of serendipity in recommender systems. *Knowledge-Based Systems, 111*, 180-192. doi:<http://dx.doi.org/10.1016/j.knosys.2016.08.014>

Mackie, J. L. (1990). *Ethics: inventing right and wrong*. London: Penguin Books.

Mitchell, T. M. (1997). *Machine Learning*. Burr Hill, IL: McGraw Hill.

Mitchell, T. M. (2006). *The discipline of machine learning* (Vol. 9): Carnegie Mellon University, School of Computer Science, Machine Learning Department.

Ngo, H., Luciw, M., Foerster, A., & Schmidhuber, J. (2012). *Learning skills from play: artificial curiosity on a katana robot arm.* Paper presented at the Proceedings of the 2012 International Joint Conference of Neural Networks, Brisbane, Australia.

Osimo, S. A., Pizarro, R., Spanlang, B., & Slater, M. (2015). Conversations between self and self as Sigmund Freud—A virtual body ownership paradigm for self counselling. *5*, 13899. doi:10.1038/srep13899

<https://www.nature.com/articles/srep13899> - supplementary-information

Pan, X., Fertleman, C., Collingwoode-William, T., Antley, A., Brenton, H., Congdon, B., . . . Delacroix, S. (2018). *A study of professional awareness under cognitive load, using immersive virtual reality: the responses of general practitioners to child safeguarding concerns'*.

Proust, M. (1996). *In search of lost time* (C. K. Scott Moncrieff & T. Kilmartin, Trans. Vol. 5). London: Vintage.

Putnam, H. (2004). *Ethics wtihout ontology*. Cambridge Mass.: Harvard University Press.

Reed, R. C. (2013). Euthyphro's Elenchus Experience: Ethical Expertise and Self-Knowledge. *Ethical theory and moral practice, 16*, 245-259.

Rini, R. (2017). Creating robots capable of moral reasoning is like parenting. *Aeon*.

Russell, S., Dewey, D., & Tegmark, M. (2015). Research priorities for robust and beneficial artificial intelligence. *AI Magazine, 36*(4), 105-114.

Scheutz, M., & Arnold, T. (2016). Against the Moral Turing Test: Accountable Design and the Moral Reasoning of Autonomous Systems. *Ethics and Information Technology, 18*(2), 103-115.

Silver, D., Huang, A., Maddison, C. J., Guez, A., Sifre, L., van den Driessche, G., . . . Hassabis, D. (2016). Mastering the game of Go with deep neural networks and tree search. *Nature, 529*(7587), 484-489. doi:10.1038/nature16961

Storck, J., Hochreiter, S., & Schmidhuber, J. (1995). Reinforcement driven information acquisition in non-deterministic environments. *Proceedings of the International Conference on Artificial Neural Networks, 2*, 159-164.

Tallis, R. (2011). *Aping Mankind: Neuromania, Darwinitis and the Misrepresentation of Humanity*. Durham: Acumen.

Turing, A. M. (1950). Computing machinery and intelligence. *Mind, 59*(236), 433-460.

Veale, M. (2017). *Progression thesis: Socio-technical approaches supporting responsible public sector machine learning.*
31